\documentclass{article}
\usepackage{wrapfig}

\usepackage{multirow}
\usepackage[table,xcdraw]{xcolor} \usepackage{colortbl}
\usepackage[table,xcdraw]{xcolor}

\usepackage{PRIMEarxiv}

\usepackage[utf8]{inputenc} % allow utf-8 input
\usepackage[T1]{fontenc}    % use 8-bit T1 fonts
\usepackage{xcolor} % For color support
\usepackage{hyperref} % For hyperlinks
\hypersetup{
    colorlinks=true,
    urlcolor=blue
}
\usepackage{url}            % simple URL typesetting
\usepackage{booktabs}       % professional-quality tables
\usepackage{amsfonts}       % blackboard math symbols
\usepackage{nicefrac}       % compact symbols for 1/2, etc.
\usepackage{microtype}      % microtypography
\usepackage{lipsum}
\usepackage{fancyhdr}       % header
\usepackage{graphicx}       % graphics
\graphicspath{{media/}}     % organize your images and other figures under media/ folder

\usepackage{amsmath}

\DeclareMathOperator*{\argmin}{arg\,min}

\title{\textbf{Regularized interpolation in 4D neural fields\\enables optimization of 3D printed geometries}}

\author{
  Christos Margadji\textsuperscript{1}, Andi Kuswoyo\textsuperscript{1}, Sebastian W. Pattinson\textsuperscript{1}
  \\
  \textsuperscript{1} Department of Engineering, University of Cambridge}
\begin{document}
\maketitle

\renewcommand{\thefootnote}{}

\footnotetext{\hspace{-1.8em} Correspondence: \texttt{\{cm2161, swp29\}@cam.ac.uk}}

\footnotetext{\hspace{-1.8em} Code: \href{https://github.com/cam-cambridge/4D-neural-fields-optimise-3D-printing}{\textcolor{magenta}{\texttt{https://github.com/cam-cambridge/4D-neural-fields-optimise-3D-printing}}}

}
\renewcommand{\thefootnote}{\arabic{footnote}}% Restore numbering

\begin{abstract}
The ability to accurately produce geometries with specified properties is perhaps the most important characteristic of a manufacturing process. 3D printing is marked by exceptional design freedom and complexity but is also prone to geometric and other defects that must be resolved for it to reach its full potential. Ultimately, this will require both astute design decisions and timely parameter adjustments to maintain stability that is challenging even with expert human operators. While machine learning is widely investigated in 3D printing, existing methods typically overlook spatial features that vary across prints and thus find it difficult to produce desired geometries. Here, we encode volumetric representations of printed parts into neural fields and apply a new regularization strategy, based on minimizing the partial derivative of the field’s output with respect to a single, non-learnable parameter. By thus encouraging small input changes to yield only small output variations, we encourage smooth interpolation between observed volumes and hence realistic geometry predictions. This framework therefore allows the extraction of “imagined” 3D shapes, revealing how a part would look if manufactured under previously unseen parameters. The resulting continuous field is used for data-driven optimization to maximize geometric fidelity between expected and produced geometries, reducing post-processing, material waste, and production costs. By optimizing process parameters dynamically, our approach enables advanced planning strategies, potentially allowing manufacturers to better realize complex and feature-rich designs.
\end{abstract}

% keywords can be removed

\section{Introduction}
Perhaps the most crucial aspect of a manufacturing process is its ability to create geometries with defined properties. This is also true of material extrusion (ME), the most prevalent additive manufacturing (AM) or 3D printing technique, which can produce highly complex and custom parts in a wide variety of materials. Its applications span areas from aerospace\cite{Pierre2023,Najmon2019,Yu2023,Klippstein2017}, to medical devices\cite{Zehao, Pattinson2017, Pattinson2019, Fu2024, Haghiashtiani2020}, and construction\cite{Hager2016, Roman}. ME relies on feedstock material in filament form which undergoes heating before being extruded through a nozzle onto the build surface in a layer-by-layer manner\cite{Ngo2018}. However, the complexity of the processes makes it prone to geometric and other defects. This is because the printed object results from the interaction of diverse factors including material flow and the motion of the mechanical system, all of which must be correctly coordinated at each point in space and time\cite{Awasthi2021}.

The complexity of the parameter space combined with the potential wealth of data available to characterize the process makes ME amenable to machine and particularly deep learning approaches. These have been widely investigated for monitoring and controlling process parameters with a view to minimizing defects\cite{Mohamed,Brion2022NatComms, Brion2022, Liu2019, Jin2019}. However, the mitigation of flaws in AM via deep learning, while always challenging, becomes increasingly intractable when compounded by sub-optimal part design that does not account for the manufacturing process\cite{ Thompson2016}. For instance, the inclusion of intricate features, such as thin walls, can introduce manufacturing instabilities that compromise geometric fidelity. In such cases, controlling material flow rate is particularly critical. Higher flow rates may be needed during initial layers to improve adhesion between the build surface and the part, while lower flow rates help maintain accuracy in thin-profile regions by preventing hatch overlap. Yet, some situations require both higher and lower flow rates within the same print, creating inevitable trade-offs. Conventionally, identifying correlations between design features and process parameters relies heavily on trial-and-error methods, which is expensive in time, labor and resources. 

Similar challenges in spatial domains have been successfully addressed using 3D deep learning. In medicine, 3D deep learning has been used to analyze breast magnetic resonance imaging scans and detect anomalous, potentially malignant geometric features\cite{Lang2023}. In robotics, it has been used for scene perception\cite{Matsuki2024} and completion\cite{Cheng2021}, allowing agents to infer occluded structures beyond their immediate observations. Additionally, it has been used to model deformations, predict structural integrity, and simulate fluid dynamics\cite{Bronstein2021, Jiang2022, Hu, Huang2020, Pfaff2021}. However, translating such methods to AM is difficult. Unlike some more established domains, AM processes often feature a combinatorial explosion of variables – driven by factors including the need to apply specific process parameters at each point in a build, properties of multiple materials and their interfaces, and highly diverse available design configurations. Complicating matters further is the scarcity of high-quality AM datasets, the difficulty of tracking defects in 3D spaces, and the limitations inherent to prevalent shape representations\cite{Wang2024}. Consequently, direct adaptation of strategies from medicine, robotics, and other domains is non-trivial, demanding new approaches tuned to the interplay between spatial and process-driven challenges of AM.

Neural fields are continuous functions (fields) – often defined over space, time, or other domains – parameterized by a neural network. Instead of storing or describing a complex object as a traditional dataset of discrete points, pixels, or voxels, a neural field represents it as an implicit function that can produce values at any coordinate. This approach allows for flexible representations that can capture intricate details and make it possible to generalize to unseen data. Neural fields have been used in various domains such as computer graphics\cite{Chen2023, Takikawa2021, Chen2021, Reddy2022, Byra, Yang2023}, robotics\cite{Wen2023, Ortiz2022, Irshad2022, Zhi2023, Mazur2023}and scientific simulations\cite{Rao2020, Jin2021, Raissi2019} and various types of fields can be modeled, including image pixel information\cite{Sitzmann2020, Stanley2007}, spectrogram features or wavelengths in audio processing tasks\cite{Sitzmann2020,Su2022,Szatkowski2024}, surrogate models for partial differential equations\cite{Raissi2019, Jiang2020}, or other relevant physical quantities\cite{Gropp2020, Simeonov2021, Chitturi2023}. Specifically for 3D representation tasks, neural fields have been utilized to represent shapes as occupancy\cite{Mescheder2019} or signed distance functions\cite{Park2019} and scenes as radiance fields\cite{Mildenhall2020}. Notable advancements include the introduction of periodic activation functions\cite{Sitzmann2020}, which capture high-frequency details in input data, and learning Fourier features directly\cite{Tancik2020}, which enhances detail representation. Other advancements have further improved the efficiency and scalability of neural fields, making them more practical for a wide range of applications\cite{Yu2021, Martel2021,Muller,Xie2022}. 

In this study we deployed neural fields to optimize geometric fidelity in 3D printing. Every neural field is designed and optimized to encode the volume of multiple instances of a geometry, as a function of the underlying process parameter. To ensure smooth interpolation, we introduce a novel gradient-driven interpolation regularization (GDIR) strategy, which minimizes the partial norm of the Jacobian of the network. This reduces the impact of small parameter variations on output geometry, drawing inspiration from the small-motion assumption and advancements in physics-informed neural networks. By treating the trained neural field as a model capable of inferring geometry at unseen flow rate values, we identify optimal process parameters to maximize geometric fidelity between expected and produced parts. Applying these parameter adjustments in the physical process demonstrates the effectiveness of our approach in enhancing print quality and fidelity.

\begin{figure}[t]
    \centering
    \includegraphics[width=1\linewidth]{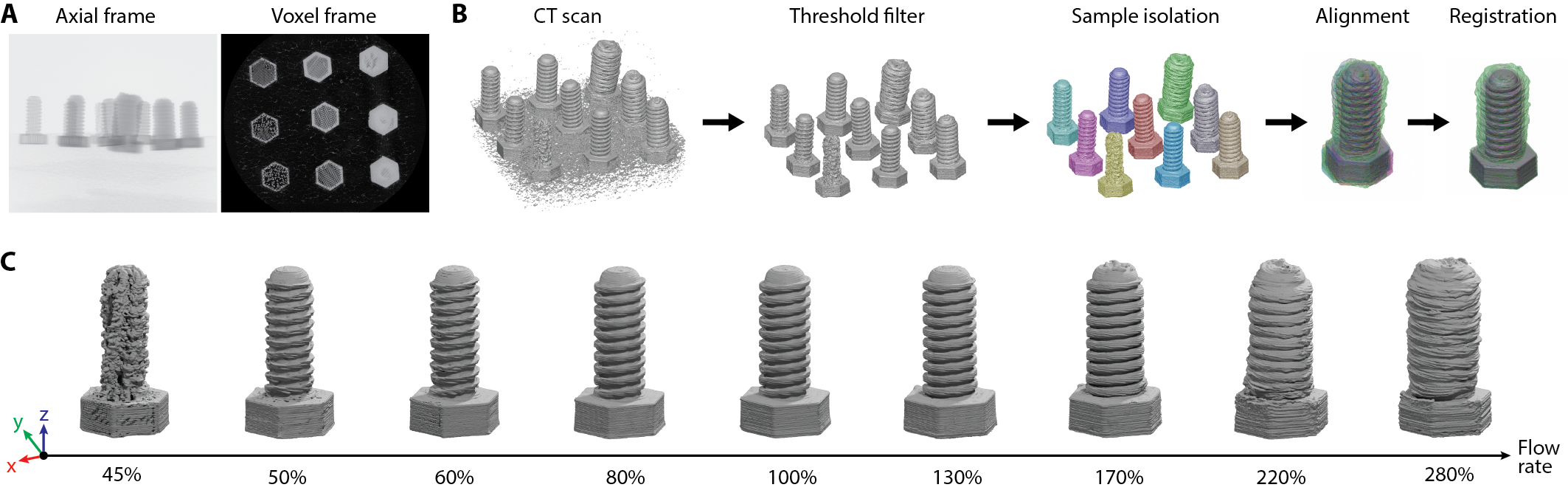}
    \caption{\textbf{Data collection and processing pipeline.} \textbf{A. }Sample axial and voxel frames obtained from the computer tomography procedure. \textbf{B. }Data processing pipeline developed for cleaning and registering different trials from the same geometry manufactured at different flow rate values. \textbf{C. }Rendered parts from the Bolt trials.}
    \label{fig:1}
\end{figure}

\section{Results}
\label{sec:headings}

\subsection{Dataset and problem formulation}
Process parameters in manufacturing, like material flow rate in 3D printing, are critical in aligning expected and produced geometries in ME. Here, "produced geometry" refers to the physical output of the process, while "expected geometry" denotes the as-modeled design specified by the user. Discrepancies between these geometries are common, which makes achieving geometric fidelity a well-documented challenge. However, datasets which systematically explore the relationship between expected and produced geometries remain scarce. To address this gap, we created a custom dataset using a 3D printing system and a computer-tomography (CT) scanner, as detailed in Methods \ref{methods:sample_prep} and \ref{methods:ct_scan}. Four experimental cycles were conducted, producing iterations of four distinct geometries: Bolt, Gear, Bunny, and Statue. Each geometry was manufactured nine times with different flow rate values (45\%, 50\%, 60\%, 80\%, 100\%, 130\%, 170\%, 220\%, and 280\%). The flow rate is expressed as a percentage relative to a calibrated baseline, where 100\% represents the optimal setting. Post-manufacturing, CT scanning was used to capture the volumetric structure of the printed parts, including internal and external features, creating digital twins of all nine iterations of each produced geometry. The scans were filtered, registered, and aligned as described in Methods \ref{methods:sample_registration}. The complete pipeline - from scan acquisition to registered volumetric data - is illustrated in Figure \ref{fig:1}B. Post-processing of each scan was validated using the physical to digital weight comparison as described in Methods \ref{methods:pdwc}.

In Figure \ref{fig:1}C, we render the processed Bolt geometry iterations. Renderings of all the digital twins from the other datasets are provided in Figure \ref{fig:S1}. Although the expected geometry is always the same (i.e., same numerical commands given to the printer), higher flow rates result in increased material deposition, leading to bulkier geometries. Conversely, at lower flow rates, insufficient material is deposited to fully fill the geometry domain, resulting in voids within the produced part. This is consistent across all tested geometries, but notably, for the Bunny geometry it is evident that specific features print more accurately at a flow rate of 170\% compared to the standard 100\%. This challenges the conventional practice of maintaining a constant flow rate throughout the process that oversimplifies the process. For instance, it might be the case that thin features may benefit from increased flow rates, due to enhanced material adhesion and improved structural integrity, compensating for the challenges of depositing material on smaller, less stable regions. This raises an important and underexplored question: can process parameters be dynamically controlled to maximize fidelity for different regions of a geometry?

\subsection{Promoting smoothness in neural fields}

\begin{figure}
    \centering
    \includegraphics[width=1\linewidth]{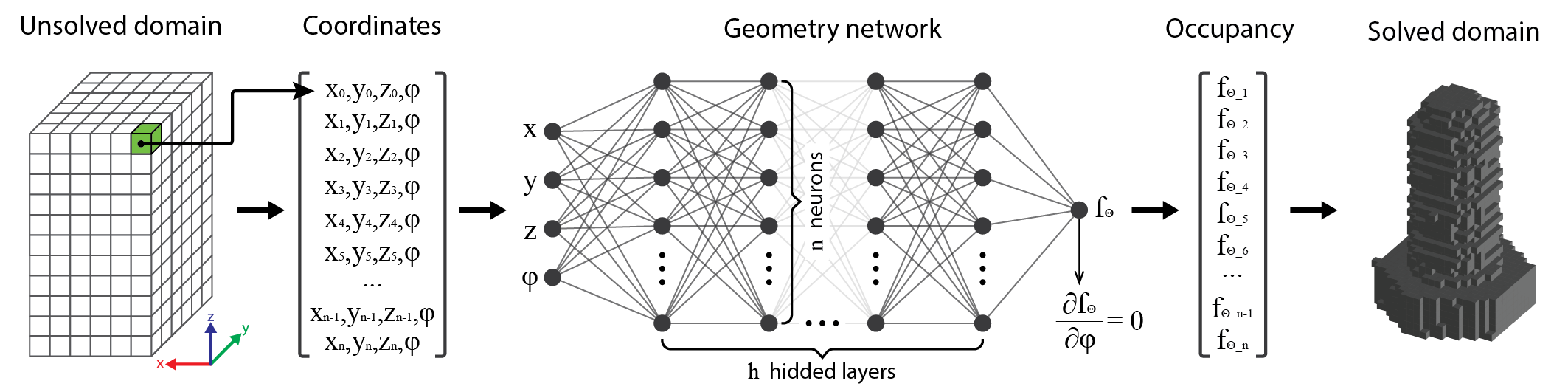}
    \caption{\textbf{End-to-end domain solver.} The domain is solved for a specific flow rate value, $\phi$, through the trained 4D neural field. The calculation of the partial derivative of the output $f_\mathrm{\Theta}$ with respect to $\phi$ and minimization of its magnitude enables smooth interpolation between seen $\phi$s.
}
    \label{fig:2}
\end{figure}

First, we aim to understand how geometry changes in response to process parameter alterations. We leverage neural fields, representing the produced geometries as learned functions denoted $F(x,y,z ;\phi$). Here, $(x,y,z)$ are spatial coordinates in 3D space, and $\phi$ represents the flow rate. The function F maps this 4D input space to a scalar value $S$, which indicates the presence ($S=1$) or absence ($S=0$) of material at the specified infinitesimal point, based on the instance's scan. In our framework, $F$ is a multi-layer perceptron featuring sinusoidal activation functions in all hidden layers but the last. To visualize the produced geometry as manufactured from a specific process parameter value $\phi=\phi_0$, we evaluate $F$ over a grid spanning the entire spatial domain, while holding $\phi_0$ constant. This allows determining the structure’s volumetric representation as shown in Figure \ref{fig:2}, referred to as “manifold reconstruction” hereafter. 

To train the neural field, we represent each CT scan as a set of data points $P$ where each point $p_i$ corresponds to a tuple $(x_i,y_i,z_i,\phi_i,S_i)$, with $(x_i,y_i,z_i)$ denoting the spatial coordinates, $\phi_i$ the process parameter,
\begin{wrapfigure}{r}{0.5\textwidth}
    \centering
    \includegraphics[width=0.5\textwidth]{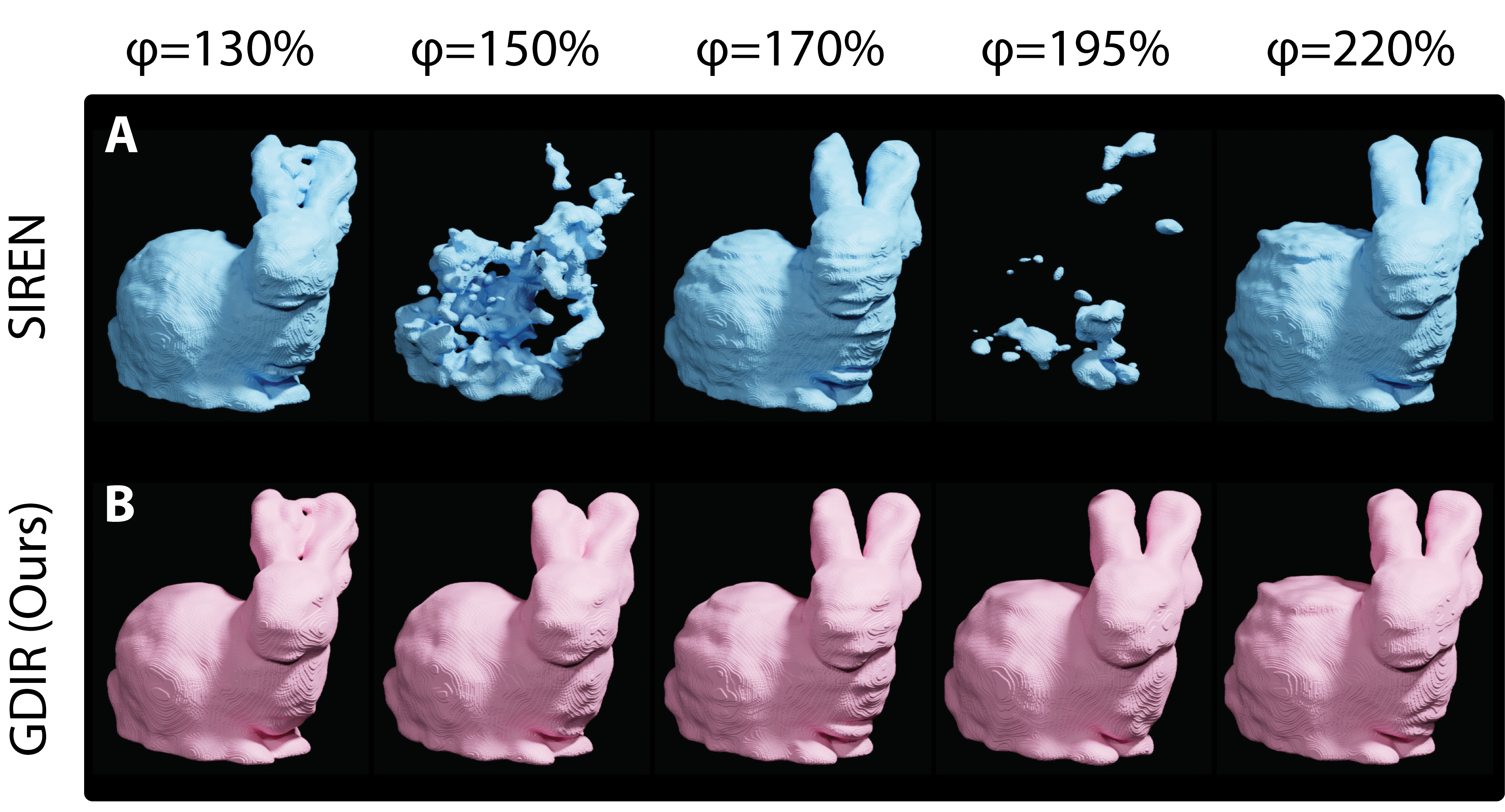}
    \caption{\textbf{Interpolation results.} While SIREN (blue) collapse at unseen regimes, our network (pink) generates smoother interpolation results.
}
    \label{fig:3}

    \vspace{5mm}

    \centering
    \includegraphics[width=0.5\textwidth]{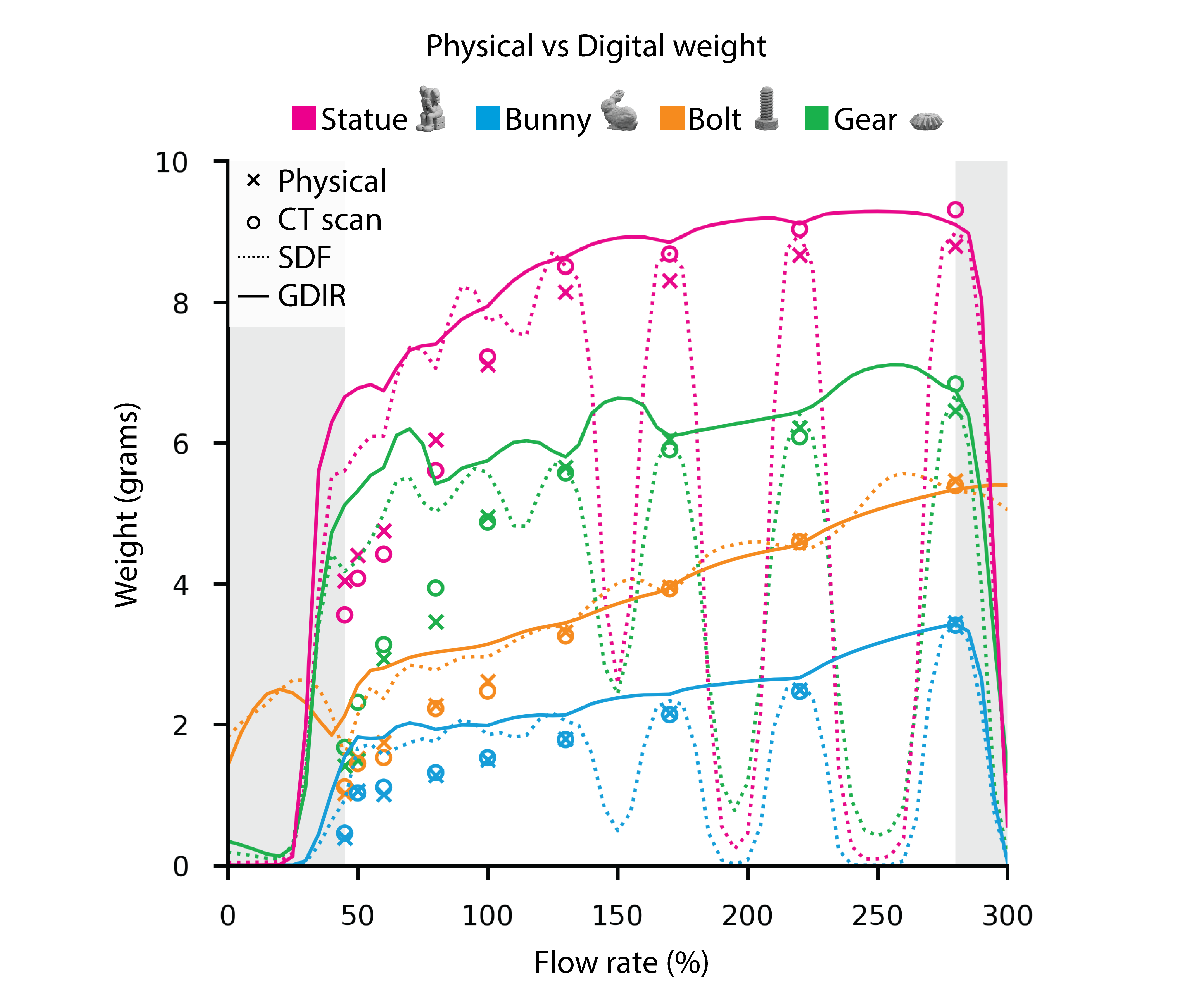}
    \caption{\textbf{Physical vs Digital weights.} We compare the physical and digital weights of the samples, reconstructed from CT scans, SIREN and our networks.}
    \label{fig:4}
\end{wrapfigure}
and $S_i\in0,1$ the material state. The number of data points N, is given by the product of the dimensions of the 3D array representing the scan resolution, multiplied by the number of available iterations per geometry. All inputs are normalized to the range $[-1,+1]$ using min-max normalization, based on the specified domain boundaries. During training the primary objective is to minimize a mean squared error function that minimizes the difference between the neural field’s prediction and the known state $S$. More training details are provided in Methods \ref{methods:vanilla} and implementation details in Methods \ref{methods:implementation}.

By training on multiple geometry instances simultaneously, we aim to construct a continuous representation of the geometric space. That would allow the learned function to ‘imagine’ how the geometry would look like, had it been manufactured under certain unseen conditions. However, there are no formal guarantees for continuity. As supervision decreases, the learning algorithm struggles, leading to a higher likelihood of unreliable interpolations between seen conditions. This phenomenon is clearly demonstrated in Figure \ref{fig:3}, where the 3D reconstructions collapse at unseen regimes, particularly in the higher ranges where supervision is more limited. Addressing this in conventional ways would require enriching the dataset with additional iterations of the geometry. Unfortunately, this approach is generally impractical due to the high cost of manufacturing and measuring the parts needed to produce such data.

The physics of the problem can be leveraged to ensure smoothness in the learned field, by encouraging Lipschitz continuity through GDIR (see Methods \ref{methods:lipschitz}) This property ensures that small changes in the input dimensions lead to proportionally small or smaller changes in the output, formally defined in equation (1):
\begin{equation}
\forall\mathcal{X}\in\mathbb{R}^3\lim_{\varphi_1 \to \varphi_2}{\left|\mathcal{F}\left(\mathcal{X},\varphi_1\right)-\mathcal{F}\left(\mathcal{X},\varphi_2\right)\right|}\approx0
\end{equation}

; where $\phi_1$ and $\phi_2$ are neighboring flow rate values and $|\cdot|$ denotes a suitable norm measuring the discrepancy between output signals. On the global scale, this implies for an infinitesimal change in the flow rate value, there should only be a negligible change in the structure’s volume as reconstructed by F. In our application we achieve this systematically, by minimizing the field’s gradients; this involves differentiating through the network with respect to the flow rate parameter input (i.e., calculate the partial derivative) and encourage smoothness at the said dimension by penalizing the norm of the gradient’s Jacobian. This becomes our secondary training objective, with effective results shown in Figure \ref{fig:3}. We also show additional results in Figure \ref{fig:S2} in random XY planes, for better visualization. We note that introducing GDIR slows down training by 8.30±0.12\%. At inference time, there is no additional latency as the gradient calculations are not necessary. 

Structural Similarity Index (SSIM) and L1-norm were employed to quantitatively evaluate the proposed method. The interpretation of these metrics and the associated error bars is provided in Methods \ref{metrics}. Metrics for seen conditions were computed by comparing reconstructed volumes with existing CT scans at previously observed flow rate values. For unseen conditions, the metrics were calculated by comparing reconstructed volumes with CT scans of additional prints produced at previously unseen flow rate values: 250\% (bolt), 115\% (bunny), 70\% (gear), and 35\% (statue). 

The results of these analyses are summarized in Table \ref{table:A1} and Table \ref{table:A2}. Under observed conditions, the smoothness objective neither clearly benefits nor hinders performance, as the empirically tuned parameter effectively balances the ability to train without disrupting gradient flow. While the standard SIREN shows slightly better performance in capturing finer details, this advantage is likely linked to its unconstrained optimization. Further experiments reveal that the performance gap between our network and SIREN narrows with additional training epochs. However, similar improvements for SIREN come at the expense of overfitting, which significantly degrades generalization in the interpolation tasks. In contrast, under unseen conditions, the superiority of the regularized networks becomes evident, as they consistently outperform standard SIRENs across all experiments and metrics.

In the same set of experiments, we compare the performance of our method when the field is encouraged to learn a signed distance function (SDF) rather than occupancy, as detailed in Methods \ref{methods:vol2sdf}. Transitioning the representation format from occupancy grids (voxels) to SDF while training with the secondary objective significantly enhances generation accuracy. The enhanced performance is attributed to the smoother gradients of SDFs, which avoid the abrupt transitions characteristic of voxel-based representations (e.g., sudden jumps from 0 to 1). This facilitates more efficient learning in the gradient space. 

The physical to digital weight comparison serves as another quantitative indicator which shows the stability of the GDIR networks in predicting the volumetric structure of a geometry at any flow rate value, Figure \ref{fig:4}. This is a useful metric for our problem, given it can exploit expected physics-intuitive trends, where the mass of an object (and thus the volume of its digital reconstruction) should be proportional to the flow rate value. We iterate through all possible flow rate values between 0\% and 300\% at intervals of 5\% and reconstruct the geometry. We then measure occupancy for each render and plot it versus flow rate. Due to the computational intensity of reconstructing the volumes needed for this graph (total of 120 volumes per part), experiments were run at half the original resolution for all geometries. Ground truth weight and CT scan weight are also included for comparison. While the GDIR method outputs an almost linear relationship between material flow rate and sample weight, SIREN exhibits a wave-like pattern, particularly in the over-extrusion regime with sparse supervision. This behavior likely stems from SIREN’s sinusoidal activations (e.g. $sin(\omega x$) which, while effective for capturing fine details, introduce periodic artifacts in poorly supervised regions. The tendency to overfit to periodic patterns under sparse data again highlights a limitation of SIREN for tasks requiring smooth, monotonic relationships. 

\subsection{Neural fields as simulators for optimizing geometric fidelity}

These models accurately represent printed objects and smoothly interpolate between seen instances; however, the primary goal is to provide a tool for engineers to maximize geometric fidelity between expected and produced geometries. To achieve this, we leverage the learned neural field as a simulation engine to model geometry evolution as a function of process changes. The network's continuity enables it to “imagine” expected outputs, even if process parameters change mid-print. This capability allows us to conduct millions of virtual experiments with varying parameter sets, facilitating dynamic discovery of optimal parameters (see Methods \ref{methods:optimisation}).

\begin{figure}
    \centering
    \includegraphics[width=1\linewidth]{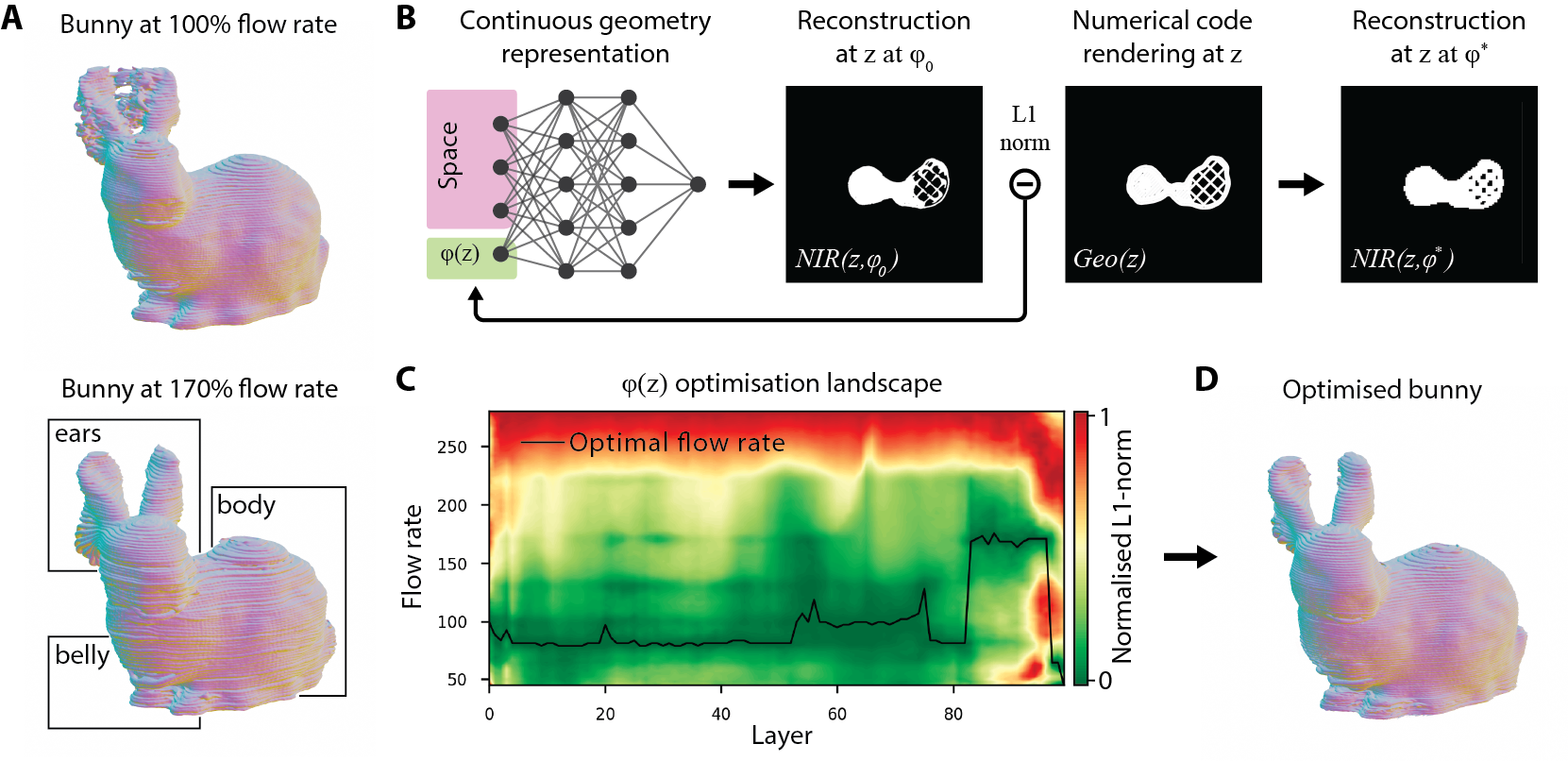}
    \caption{\textbf{Geometric fidelity optimization results.} \textbf{A.} Renders of the produced geometry of the Stanford bunny as manufactured at 100\% and 170\% flow rate values. \textbf{B.} The optimization pipeline compares the learned field reconstructions with renderings from the expected geometry. \textbf{C.} The optimization landscape. For every layer of the print, the optimal flow rate value $\phi$* is the selected such that reconstructed and expected geometries match. \textbf{D.} Applying the values on a per-layer basis yields better geometric outcomes.
}
    \label{fig:5}
\end{figure}

To evaluate our methodology, we optimized the geometric fidelity of the printed Bunny model, a widely used yet challenging case study due to its intricate features. Our goal was to determine the optimal flow rate for each of the model's 100 layers to preserve smoothness in the main body and minimize overlap in delicate regions like the ears (Figure \ref{fig:5}A). Treating flow rate as a discrete variable ranging from 45\% to 280\% in 1\% increments, we computed a fitness score proportional to the L1-norm between the reconstruction and the rendering (Figure \ref{fig:1}B). This produced a grid landscape where the optimizer sought the minimum trough corresponding to the optimal parameters (Figure \ref{fig:1}C).

The optimizer identified higher flow rates for the ear regions while reducing material extrusion early to prevent excessive buildup at the ear tips. It also slightly under-extruded the body, a common 3D printing practice to enhance geometric fidelity, albeit at the expense of structural integrity. Applying these corrections resulted in a printed Bunny with both smooth body contours and fine ear details, Figure \ref{fig:1}D. However, the unsupported belly area remained rough, as the optimization algorithm treated each layer independently. This limitation prevented it from addressing roughness caused by insufficient support from preceding layers in the 3D space.

\section{Discussion}

Central to our approach is a novel regularization strategy that promotes smooth interpolation across the process parameter space. By introducing structured inductive biases in regions where data is sparse, this technique helps avoid erratic behavior in uncharted areas. In essence, it guides the model toward more stable and predictable outputs, even when direct supervision is limited. Most importantly, our approach is process and representation agnostic, which implies that it may be used for any other manufacturing method which requires optimization to achieve diverse properties as a function of process parameters. Additionally, we note this approach is well applicable to a broader set of problems within the machine learning community, including video frame interpolation.

The shapes of produced parts from 3D printing are a function of the expected geometry, and the process parameters used to produce it as we showed. We empirically learn this function, utilizing neural fields as a form of continuous shape representation. The neural fields are directly learned from experimental data obtained through CT scans of the manufactured parts and gradient regularization is deployed to smoothen them. The trained networks serve as a physics-intuitive tool use to extract accurate volume reconstructions of objects printed at previously unseen process parameters. Identification of optimal process parameters on a per-layer basis is thus enabled, to maximize geometric fidelity between expected and produced geometry. By applying the suggested changes in a dynamic manner, we can produce parts with unprecedented accuracy.

Micro-CT scanning provides a comprehensive digital twin of produced parts, capturing both external and internal features. However, it may be impractical on factory floors due to complexity and cost. As a simpler alternative, structured light scanning can capture external geometries at a lower cost, enabling the use of straightforward loss functions (e.g., those leveraging surface normals) and focusing on aesthetics and dimensional tolerances. Although structured light lacks internal data, its robustness and reduced noise often suffice for many industrial needs. In this work, we demonstrate our approach using micro-CT data, which is the more complex scenario. Consequently, if it can handle the intricacies of internal and external features derived from micro-CT scans, it will readily apply to the simpler, external-only case offered by structured light scanning. This ensures flexibility across different scanning technologies and industrial contexts.

In summary, this work presents a way to predict geometric and local properties of manufactured parts, enabling both autonomous quality control and process parameter optimization, paving the way for advanced manufacturing strategies that move beyond one-size-fits-all settings. While the initial datasets can be laborious to construct, the methodology coupled to these efficient processing pipelines may allow for significantly better outcomes across various complex and feature-rich designs. More broadly, this work lays the foundation for wider research into neural fields capable of optimizing process parameters for unseen geometries in a one-shot manner. 

\newpage
\section{Methods}

\subsection{Experimental design}

\subsubsection{Sample preparation}
\label{methods:sample_prep}
The experimental setup for data collection consisted of a Creality CR-20 Pro material extrusion AM system, without additional modifications to replicate typical operational conditions. The AM system was interfaced with a Raspberry Pi 4 Model B running OctoPrint 1.9.3. The material flow rate, defined as the percentage of material exiting the nozzle's orifice per unit of time, was directly controlled by passing numerical commands (M221 SXX where XX is given in percentage) to the printer's firmware. This measurement is always relative to the default settings established during calibration. The AM system utilized polylactic acid (PLA) material feedstock in the form of a 1.75 mm grey color filament, sourced from PolyMaker.

All geometries were outsourced from the internet in a mesh form and were turned into GCode using Cura 5.6.0. The printing durations for each trial were P01: 47 minutes, P02: 68 minutes, P03: 23 minutes and P04: 32 minutes. Each geometry was manufactured nine times, each time with a different flow rate as explained in the main text. The parts from each geometry were then scanned using computer tomography and the processing pipeline described in the main text was used for alignment and registration. All data are shown in Supplementary Figure \ref{fig:S1} A to D. Physical to digital weight for the geometries is also shown in Figure S1E.

\subsubsection{Micro-Computer-Tomography}
\label{methods:ct_scan}

A Nikon Xtex 225 H micro-CT scanner with a 28802880-pixel detector was used to capture the volumetric structure of the geometries. CT was chosen over alternative methods like structured light scans due to its capability to capture internal structures in addition to external surfaces without significant cost constraints. For each geometry, the nine samples were loaded on a custom foam holder and scanned together to increase efficiency. The voxel size in X, Y and Z was set to 35.126 to 45.448 microns, depending on the size of the holder. The CT scan originally provided axial images, which were transitioned to voxel data via standard computational reconstruction techniques, Figure \ref{fig:1}A. A threshold filter was then applied to eliminate lower refraction intensities occurring from the foam holder, and a Gaussian filter was applied to remove any remaining artifacts. The scan was subsequently partitioned to isolate the nine samples, and each one was moved to its own axes system.

\subsubsection{Registration and alignment}
\label{methods:sample_registration}

Orientation mismatches in CT scan analysis and other spatial domain applications often hinder the direct comparability of data from different samples. To address this challenge, we detect the first layer of each sample for precise centering and alignment. Specifically, we track the first activated pixel from the bottom along the Z axis to construct a depth map that accurately represents the bottom surface of each build. We then fit a plane on the detected depth points and calculate a transformation that aligns with the XY plane. This transformation is subsequently applied to the entire domain, ensuring that the Z direction of the part is aligned with the original build direction. We then register the different sample scans, by calculating a transformation that involves translation in the X and Y directions and rotation around the Z axis, utilizing a probabilistic approach known as coherent point drift (CPD).

\subsection{Model training}
\subsubsection{Neural fields}
\label{methods:vanilla}

In principle, a neural field is a neural function which encodes the mapping from input coordinates to corresponding signal values, Equation (2)
\begin{equation}
\Theta^\ast = \argmin_{\Theta} \frac{1}{N}\sum_{i=0}^{N}\mathcal{L}\Bigl(S_i,\,\mathcal{F}_\Theta(\mathcal{X}_i)\Bigr)
\end{equation}

; where $R^{in}$ represents the input space and each dimension corresponds to a feature or attribute of the data, $\Theta$ the trainable parameters that govern the behavior of the neural network and $R^{out}$ signifies the output space. To train the neural field, a dataset composed of N input coordinates $X_i\in R^{in}$ and corresponding output signal values $S_i\in R^{out}$  is exploited, where i indexes point pairs and takes values $1\leq i\leq N$. The training objective is finding the optimal parameters $\Theta^*$ by solving the minimization problem expressed in Equation (3)
\begin{equation}
    \mathrm{\Theta}^\ast={\rm argmin}\mathrm{\Theta}{\frac{1}{N}\sum_{i=0}^{N}\mathcal{L}\left(S_i,\mathcal{F}_\mathrm{\Theta}\left(\ \mathcal{X}_i\right)\right)}
\end{equation}

; where $L$ is a point-wise loss function that quantifies the difference between the predictions $F_\Theta(X_i)$ and the ground truth values $S_i$, typically implemented as a mean squared error (MSE) function, Equation (4)
\begin{equation}
    \mathcal{L}_{MSE}=\frac{1}{N}\sum_{i=0}^{N}\left(S_i-\mathcal{F}_\mathrm{\Theta}\left(\ \mathcal{X}_i\right)\right)^2
\end{equation}

\subsubsection{Gradient-driven interpolation regularization}
\label{methods:lipschitz}

A Lipschitz network is a type of neural network designed to enforce a Lipschitz constraint, which ensures that the rate of change of the network's output is bounded by a constant c, referred to as the Lipschitz constant\cite{Liu2022}. Formally, for any two inputs $\phi_1$ and $\phi_2$ in the input space $R^{in}$, a Lipschitz continuous function $F_\Theta$ satisfies Equation (5)
\begin{equation}
\left|\left|\mathcal{F}_\mathrm{\Theta}\left(\varphi_1\right)-\mathcal{F}_\mathrm{\Theta}\left(\varphi_2\right)\right|\right|\le\ c\left|\left|\varphi_1-\varphi_2\right|\right|
\end{equation}

; where c controls the smoothness of the function. This property is critical for tasks requiring stable, controlled, and interpretable mappings, such as neural fields, where the network encodes spatially continuous functions like signed distance fields. Similar assumptions have been made elsewhere\cite{Drucker1991, Moosavi, Elsner2021} including for optical flow filters which are famous for motion prediction\cite{Lyasheva2020}.

Without losing generality, Equation (5) can also be expressed as Equation (1) from the main text, which can be enforced by encouraging the partial derivative of the network's output with respect to $\phi$ being equal to 0, Equation (6)
\begin{equation}
    \frac{\partial\mathcal{F}_\mathrm{\Theta}}{\partial\varphi}=0
\end{equation}

The partial derivative can be computed using automatic differentiation\cite{Paszke2017} which is readily available in most common machine learning frameworks such as autograd in PyTorch. To encourage minimization of the computed derivative, an additional loss term is integrated, with the loss function becoming Equation (7)
\begin{equation}
\mathcal{L}=\mathcal{L}_{MSE}+\lambda\cdot\left|\frac{\partial\mathcal{F}_\mathrm{\Theta}}{\partial\varphi}\right|_2^2
\end{equation}

; where $\lambda$ is a weight factor to balance the magnitudes of the two terms. This factor controls the strength with which gradient constraints are enforced, leading to a constrain-malleability trade-off as discussed. We call this approach gradient-driven interpolation regularization (GDIR) as it promotes smoothness in the learned representation by regulating its gradients.

We note that the minimization of the two loss terms is computed on two different point sets. While $L_{MSE}$ is computed on the available data points which represent occupancy (from the CT scans), the minimization of the partial derivative needs to hold true at all coordinates in $R^{in}$. We thus utilize Latin hypercube sampling (LHS) to generate proxies for the latter, i.e., random coordinates across the combined spatial and parameter spaces. LHS is particularly advantageous as it ensures a more uniform distribution of sampled points across the entire space compared to simple random sampling. The efficiency and accuracy of this method also depend on the number of sampled proxies. Generally, a higher number enhances the resolution and robustness of GDIR at the cost of increased computational intensity. It is crucial to balance the granularity of sampling with the available computational resources.

\subsection{Volume to signed distance function}
\label{methods:vol2sdf}
We convert the volumetric data into a signed distance function (SDF) by applying the marching cubes algorithm\cite{Lorensen1987} to extract surfaces from each available volume. We then calculate the Euclidean distance of each point in the volumetric grid from its nearest surface, resulting in the SDF. This scalar field encodes the distance from each point to the nearest surface, with the sign indicating whether the point is inside or outside the geometry. Specifically, points inside the geometry have negative SDF values, points on the surface have a zero SDF value (the zero-level set), and points outside have positive SDF values. The zero-level set precisely defines the object's surface. All SDF values are normalized based on their minimum and maximum values according to their sign. The grid is then flattened into $N$ points with form $(x,y,z,\phi,SDF)$.

\subsubsection{Implementation details}
\label{methods:implementation}
The architecture we use is based on the original SIREN with hyperparameters shown in Table \ref{table:A3}. GDIR networks have same characteristics, but the additional regularization term is enabled. For each geometry, the best frequency for the sinusoidal activations (both first and hidden) was found using a grid search on 10\% of the available training data points. Training and inference were conducted on one NVIDIA Quadro RTX 5000 (16GB) GPU supplemented by an Intel Core i9-9900K (16-core) CPU and 64GB of DDR4 RAM.

\subsection{Optimization strategy}
\label{methods:optimisation}

Assuming the neural field achieves high accuracy, we iteratively evaluate all layers to determine the optimal flow rate $\phi$ that maximizes the similarity between the cross-section of the expected geometry at any given height z and the corresponding manifold reconstruction. This approach is conceptually like topology optimization methods (also solvable with neural fields)\cite{Zehnder2021, Nobari2024} but focuses on precise control of material deposition at a per-layer resolution to achieve desired geometric outcomes. This can be expressed as Equation (8)
\begin{equation}
    \varphi\left(z\right)={\rm\argmin_{\varphi}{\left|\mathcal{F}_\mathrm{\Theta}\left(x,y\middle|\ z\right)-layer\left(z\right)\right|}}
\end{equation}

; where $layer(z)$ represents cross-sectional images rendered from the expected geometry, and $F_\Theta(x,y|z)$ corresponds to the geometric reconstruction generated by the field at the pre-specified height $z$. The "rendered" cross-sections represent the expected geometry and are obtained by processing numerical outputs from the computer-aided manufacturing slicer and visualizing them using computer graphics techniques, rather than directly using surfaces derived from the CAD mesh. This ensures fidelity not only to external surfaces but also to internal features influenced by infill patterns, which are critical for the part's functionality.

\subsection{Validation metrics}
\label{metrics}

\subsubsection{Physical to digital weight comparison}
\label{methods:pdwc}

To calculate the digital weight and compare it with the physical weight, we calculate the product of occupied voxels, volume of each voxel as a function of resolution, and material density. In mathematical form, Equation (9)
\begin{equation}
    DW=O\times\ v\times\rho
\end{equation}

; where $O$ is a scalar representing the occupancy in terms of the number of activated voxels, $v$ is the volume of each voxel and $\rho$ is the density of polylactic acid (PLA) equal to $1.25 g/cm^3$. The physical mass of the samples is weighted using an Ohaus Navigator NVT Precision Balance.

\subsubsection{Structural Similarity Index}

The Structural Similarity Index (SSIM) evaluates the similarity between two images, emphasizing structural content over pixel-wise differences. Widely used in tasks such as compression, denoising, and image generation, SSIM compares luminance (mean intensity), contrast (standard deviation), and structure (normalized correlation) across small, overlapping windows, typically weighted by a Gaussian function. It is defined as: Equation (10)
\begin{equation}
SSIM\left(x,y\right)=\frac{\left(2\mu_x\mu_y+C_1\right)\left(2\sigma_{xy}+C_2\right)}{\left(\mu_x^2+\mu_y^2+C_1\right)\left(\sigma_x^2+\sigma_y^2+C_2\right)}\ 
\end{equation}

; where $\mu_x$ and $\mu_y$ are the mean intensities, $\sigma_x$ and $\sigma_y$ are the variances, and $\sigma_xy$ is the covariance between images $x$ and $y$. Constants $C_1$ and $C_2$ stabilize the formula to avoid division by zero. SSIM values range from $[-1,+1]$, where the max value indicates identical images.

For our 3D domain, we compute SSIM by slicing it into 2D cross-sections along the Z-axis, calculating SSIM for each slice. Results are averaged, with error bars representing standard deviations across slices.

\subsubsection{L1-norm}

The L1-norm is a suitable choice for comparing discrepancies between two volumetric datasets because it measures the absolute differences between corresponding elements in the datasets. By summing the absolute values of these differences, the L1-norm provides a straightforward and interpretable metric to quantify the overall deviation between the datasets, making it ideal for evaluating similarity or identifying anomalies in volumetric data.

Mathematically, for a pair of volumetric datasets with the same resolution, the L1-norm is defined as Equation (11):
\begin{equation}
    \left|V_1-V_2\right|_1=\frac{1}{N}\sum_{i=1}^{n}\sum_{j=1}^{m}\sum_{k=1}^{p}\left|V_1\left(i,j,k\right)-V_2\left(i,j,k\right)\right|
\end{equation}

; where $V_1(i,j,k)$ and $V_2(i,j,k)$ are the values of the volumetric datasets $V_1$ and $V_2$ at the spatial coordinate $(i,j,k)$ and $n$, $m$, and $p$ are the dimensions of the datasets along the x-, y-, and z-axes, respectively. $N$ is the total number of points being compared, equal to $n\times m\times p$. This formulation provides a global measure of discrepancy, making it an effective tool for volumetric data analysis. L1-norm is averaged on across parts.

\section*{Acknowledgments}

We acknowledge the support of Engineering and Physical Sciences Research Council grant EP/V062123/1. CM was funded by the EPSRC Studentship (EP/N509620/1). AK is supported by the Jardine Foundation.  

%Bibliography
\bibliographystyle{unsrt}  
\bibliography{references}  

\clearpage
\section*{Appendix}
This document includes appendix tables and figures.
\clearpage

% Reset figure numbering for Supplementary
\setcounter{figure}{0}
\renewcommand{\thefigure}{A\arabic{figure}}

% Reset table numbering for Supplementary
\setcounter{table}{0}
\renewcommand{\thetable}{A\arabic{table}}

\begin{figure}[h]
\centering
\includegraphics[width=1\linewidth]{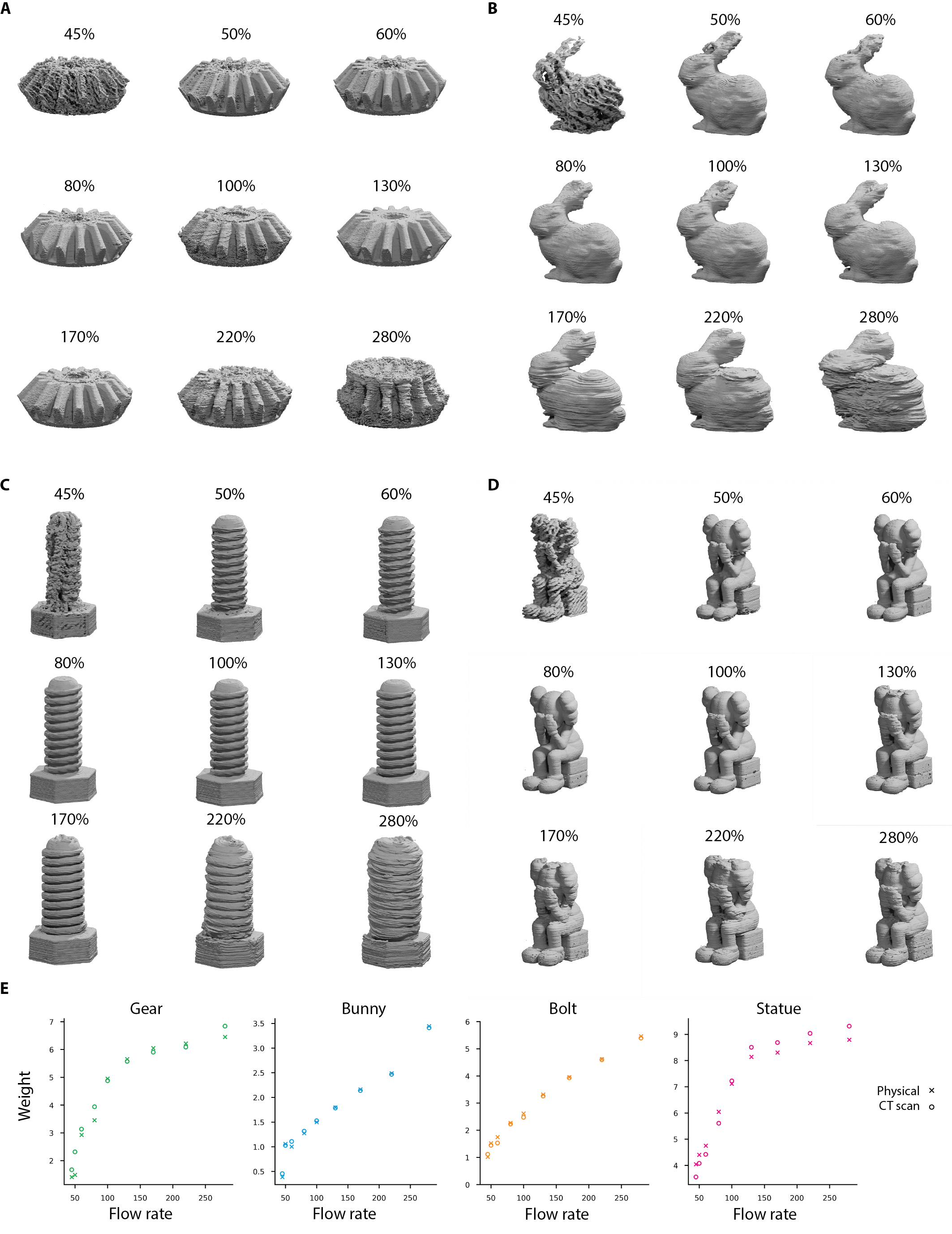}
\caption{An example supplementary figure.}
\label{fig:S1}
\end{figure}

\begin{figure}[h]
\centering
\includegraphics[width=1\linewidth]{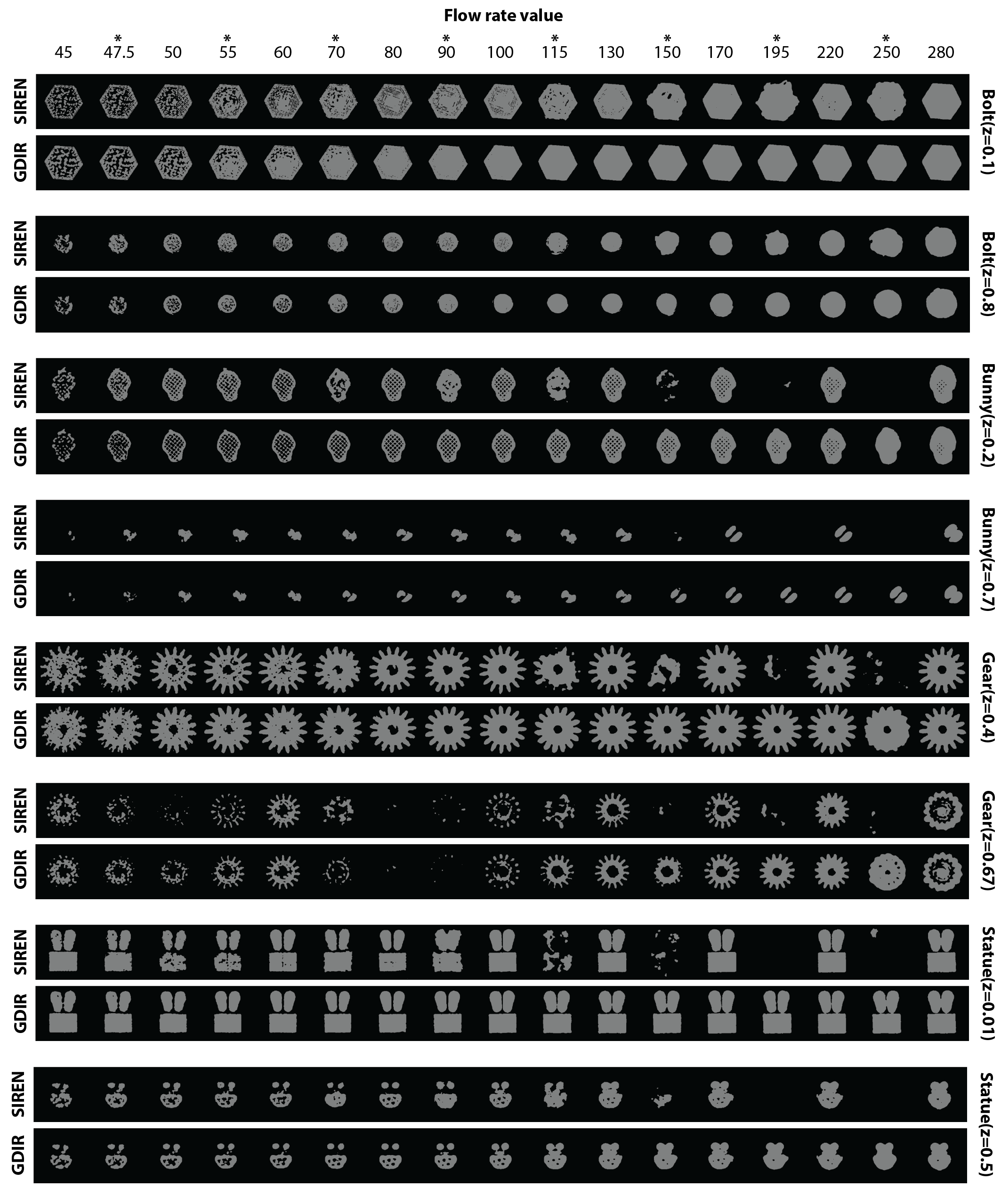}
\caption{An example supplementary figure.}
\label{fig:S2}
\end{figure}

\begin{table}[t]
\caption{Quantitative comparison of the geometric fidelity of represented shapes at seen
flowrate values.}
\renewcommand{\arraystretch}{1.5}
\centering
\begin{tabular}{ccccccccc}
\rowcolor[HTML]{A6A6A6} 
\cellcolor[HTML]{A6A6A6}                                 & \multicolumn{2}{c}{\cellcolor[HTML]{A6A6A6}\textbf{Bolt}}                                                                      & \multicolumn{2}{c}{\cellcolor[HTML]{A6A6A6}\textbf{Gear}}                                                                      & \multicolumn{2}{c}{\cellcolor[HTML]{A6A6A6}\textbf{Bunny}}                                                                     & \multicolumn{2}{c}{\cellcolor[HTML]{A6A6A6}\textbf{Statue}}                                                                    \\
\rowcolor[HTML]{D9D9D9} 
\multirow{-2}{*}{\cellcolor[HTML]{A6A6A6}\textbf{Model}} & \textbf{L1 ↓}                                                  & \textbf{SSIM ↑}                                               & \textbf{L1 ↓}                                                  & \textbf{SSIM ↑}                                               & \textbf{L1 ↓}                                                  & \textbf{SSIM ↑}                                               & \textbf{L1 ↓}                                                  & \textbf{SSIM ↑}                                               \\
\cellcolor[HTML]{D9D9D9}Siren (Occupancy)                & \begin{tabular}[c]{@{}c@{}}17.94\\ ±8.05\end{tabular}          & \begin{tabular}[c]{@{}c@{}}0.90\\ ±0.02\end{tabular}          & \textbf{\begin{tabular}[c]{@{}c@{}}22.51\\ ±9.18\end{tabular}} & \textbf{\begin{tabular}[c]{@{}c@{}}0.76\\ ±0.05\end{tabular}} & \textbf{\begin{tabular}[c]{@{}c@{}}13.49\\ ±3.47\end{tabular}} & \textbf{\begin{tabular}[c]{@{}c@{}}0.92\\ ±0.00\end{tabular}} & \begin{tabular}[c]{@{}c@{}}27.50\\ ±5.56\end{tabular}          & \textbf{\begin{tabular}[c]{@{}c@{}}0.89\\ ±0.00\end{tabular}} \\
\cellcolor[HTML]{D9D9D9}Ours (Occupancy)                 & \begin{tabular}[c]{@{}c@{}}17.81\\ ±8.18\end{tabular}          & \begin{tabular}[c]{@{}c@{}}0.89\\ ±0.02\end{tabular}          & \begin{tabular}[c]{@{}c@{}}23.05\\ ±9.82\end{tabular}          & \begin{tabular}[c]{@{}c@{}}0.76\\ ±0.05\end{tabular}          & \begin{tabular}[c]{@{}c@{}}13.76\\ ±3.40\end{tabular}          & \begin{tabular}[c]{@{}c@{}}0.92\\ ±0.00\end{tabular}          & \textbf{\begin{tabular}[c]{@{}c@{}}26.57\\ ±5.00\end{tabular}} & \begin{tabular}[c]{@{}c@{}}0.89\\ ±0.01\end{tabular}          \\
\cellcolor[HTML]{D9D9D9}Siren (SDF)                      & \textbf{\begin{tabular}[c]{@{}c@{}}16.17\\ ±8.76\end{tabular}} & \textbf{\begin{tabular}[c]{@{}c@{}}0.92\\ ±0.02\end{tabular}} & \begin{tabular}[c]{@{}c@{}}24.15\\ ±12.53\end{tabular}         & \begin{tabular}[c]{@{}c@{}}0.76\\ ±0.06\end{tabular}          & \begin{tabular}[c]{@{}c@{}}15.04\\ ±4.71\end{tabular}          & \begin{tabular}[c]{@{}c@{}}0.92\\ ±0.01\end{tabular}          & \begin{tabular}[c]{@{}c@{}}31.96\\ ±7.99\end{tabular}          & \begin{tabular}[c]{@{}c@{}}0.87\\ ±0.00\end{tabular}          \\
\cellcolor[HTML]{D9D9D9}Ours (SDF)                       & \begin{tabular}[c]{@{}c@{}}21.95\\ ±13.27\end{tabular}         & \begin{tabular}[c]{@{}c@{}}0.89\\ ±0.03\end{tabular}          & \begin{tabular}[c]{@{}c@{}}29.59\\ ±17.44\end{tabular}         & \begin{tabular}[c]{@{}c@{}}0.73\\ ±0.07\end{tabular}          & \begin{tabular}[c]{@{}c@{}}20.50\\ ±8.70\end{tabular}          & \begin{tabular}[c]{@{}c@{}}0.90\\ ±0.01\end{tabular}          & \begin{tabular}[c]{@{}c@{}}29.92\\ ±14.63\end{tabular}         & \begin{tabular}[c]{@{}c@{}}0.85\\ ±0.01\end{tabular}         
\end{tabular}
\label{table:A1}
\end{table}

\begin{table}[t]
\centering
\caption{Quantitative comparison of the geometric fidelity of represented shapes at unseen flowrate values.}
\renewcommand{\arraystretch}{1.5}
\begin{tabular}{ccccccccc}
\rowcolor[HTML]{A6A6A6} 
\cellcolor[HTML]{A6A6A6}                                                           & \multicolumn{2}{c}{\cellcolor[HTML]{A6A6A6}\textbf{\begin{tabular}[c]{@{}c@{}}Bolt\\ ()\end{tabular}}} & \multicolumn{2}{c}{\cellcolor[HTML]{A6A6A6}\textbf{\begin{tabular}[c]{@{}c@{}}Gear\\ ()\end{tabular}}} & \multicolumn{2}{c}{\cellcolor[HTML]{A6A6A6}\textbf{\begin{tabular}[c]{@{}c@{}}Bunny\\ ()\end{tabular}}} & \multicolumn{2}{c}{\cellcolor[HTML]{A6A6A6}\textbf{\begin{tabular}[c]{@{}c@{}}Statue\\ ()\end{tabular}}} \\
\rowcolor[HTML]{D9D9D9} 
\multirow{-2}{*}{\cellcolor[HTML]{A6A6A6}\textbf{Model}}                           & \textbf{L1 ↓}              & \textbf{SSIM ↑}                                                           & \textbf{L1 ↓}              & \textbf{SSIM ↑}                                                           & \textbf{L1 ↓}               & \textbf{SSIM ↑}                                                           & \textbf{L1 ↓}               & \textbf{SSIM ↑}                                                            \\
\cellcolor[HTML]{D9D9D9}Siren (Occupancy)                                          & 28.90                      & \begin{tabular}[c]{@{}c@{}}0.80\\ ±0.02\end{tabular}                      & 49.96                      & \textbf{\begin{tabular}[c]{@{}c@{}}0.58\\ ±0.03\end{tabular}}             & 34.65                       & \begin{tabular}[c]{@{}c@{}}0.85\\ ±0.05\end{tabular}                      & 80.92                       & \begin{tabular}[c]{@{}c@{}}0.67\\ ±0.07\end{tabular}                       \\
\cellcolor[HTML]{D9D9D9}\begin{tabular}[c]{@{}c@{}}Siren (SDF)\end{tabular}      & 65.34                      & \begin{tabular}[c]{@{}c@{}}0.75\\ ±0.03\end{tabular}                      & 56.74                      & \begin{tabular}[c]{@{}c@{}}0.56\\ ±0.03\end{tabular}                      & 44.57                       & \begin{tabular}[c]{@{}c@{}}0.84\\ ±0.04\end{tabular}                      & 102.05                      & \begin{tabular}[c]{@{}c@{}}0.66\\ ±0.06\end{tabular}                       \\
\cellcolor[HTML]{D9D9D9}\begin{tabular}[c]{@{}c@{}}Ours (Occupancy)\end{tabular} & \textbf{23.71}             & \begin{tabular}[c]{@{}c@{}}0.80\\ ±0.04\end{tabular}                      & \textbf{41.21}             & \begin{tabular}[c]{@{}c@{}}0.56\\ ±0.02\end{tabular}                      & 32.47                       & \begin{tabular}[c]{@{}c@{}}0.86\\ ±0.06\end{tabular}                      & 52.60                       & \begin{tabular}[c]{@{}c@{}}0.70\\ ±0.09\end{tabular}                       \\
\cellcolor[HTML]{D9D9D9}\begin{tabular}[c]{@{}c@{}}Ours (SDF)\end{tabular}       & 24.55                      & \textbf{\begin{tabular}[c]{@{}c@{}}0.84\\ ±0.03\end{tabular}}             & 46.24                      & \begin{tabular}[c]{@{}c@{}}0.58\\ ±0.03\end{tabular}                      & \textbf{26.52}              & \textbf{\begin{tabular}[c]{@{}c@{}}0.89\\ ±0.04\end{tabular}}             & \textbf{47.33}              & \textbf{\begin{tabular}[c]{@{}c@{}}0.78\\ ±0.05\end{tabular}}             
\end{tabular}
\label{table:A2}
\end{table}

\begin{table}[t]
\centering
\caption{Key training hyperparameters.}
\renewcommand{\arraystretch}{1.5}
\begin{tabular}{
>{\columncolor[HTML]{D9D9D9}}l c}
Hidden layers           & 8                               \\
Neurons per layer       & 1024                            \\
Epochs                  & 5                               \\
Batch size              & 50,000                          \\
GDIR proxies            & 50,000                          \\
Optimizer               & Rprop                           \\
Learning rate           & 1e-4                            \\
Learning rate scheduler & Cosine annealing (/10 per step)
\end{tabular}
\label{table:A3}
\end{table}

\end{document}